\documentclass[prl,aps,twocolumn,showpacs]{revtex4}
\usepackage{graphicx}
\usepackage{dcolumn}
\usepackage{bm}

\begin{document}

\title{Spinor dipolar Bose-Einstein condensates; Classical spin approach}

\author{M. Takahashi$^1$, Sankalpa Ghosh$^{1,2}$, T. Mizushima$^1$, 
K. Machida$^1$}

\affiliation{$^1$Department of Physics, Okayama University,
Okayama 700-8530, Japan}

\affiliation{$^2$Department of Physics,
Indian Institute of Technology, Delhi,
Hauz Khas,
New Delhi 110016,
India}

\date{\today}

\begin{abstract}
Magnetic dipole-dipole interaction dominated Bose-Einstein condensates are
discussed under spinful situations. We treat the spin degrees of freedom
as a classical spin vector, approaching from large spin limit
to obtain an effective minimal Hamiltonian; a version extended from a non-linear
sigma model. By solving the Gross-Pitaevskii equation
we find several novel spin textures where the mass density and spin density are
strongly coupled, depending upon trap geometries due to the long-range and 
anisotropic natures of the dipole-dipole interaction.
\end{abstract}

\maketitle

Bose-Einstein condensates (BEC) with internal degrees of freedom, 
the so-called spinor BEC have attract much attention experimentally 
and theoretically in recent years \cite{review}. Spinor BEC opens up a new
paradigm where the order parameter of condensates 
is described by a multi-component vector \cite{ohmi,ho}. 
This can be possible by optically trapping cold atoms
where all hyperfine states are liberated,
while magnetic trapping freezes its freedom.
So far $^{23}$Na (the hyperfine state $F = 1$), and  $^{87}$Rb  ($F = 2$) are 
extensively investigated.

Griesmaier {\it et al.} \cite{griesmaier} have recently 
succeeded in achieving BEC of $^{52}$Cr atom gases 
whose magnetic moment per atom is 3 $\mu_B$ (Bohr magneton). 
There has been already emerging \cite{stuhler} several novel aspects associated with 
larger magnetic moment in $^{52}$Cr atom even 
in this magnetic trapping, where all spin moments are polarized along an external
magnetic field. Namely 
the magnetic dipole-dipole (d-d)  interaction, which is proportional to $F^2$ is 
expected to play an important role in a larger spin atom.

It is natural to expect realization of BEC with still larger spin atomic species 
under the spinful situations by optical trapping or control the 
d-d interaction via the Feshbach resonance relative 
to other interaction channels. There has already been existing 
a large amount of theoretical studies for dipolar BEC \cite{baranov}. 
Most of them treat the polarized case where the dipolar moments are aligned 
along an external field. The intrinsic anisotropic or tensorial nature of the 
d-d interaction relative to the polarization axis 
manifests itself in various properties. The head-to-tail moment arrangement 
due to the d-d interaction is susceptible to a shape instability by concentrating atoms 
in the central region. We have seen already that tensorial and long-ranged d-d interaction
is responsible for this kind of shape dependent phenomenon
where the mass density is constrained  by the polarization axis.

 In contrast the theoretical studies of the 
spinor dipolar BEC are scarce, and  just started with several 
impressive works \cite{kawaguchi,pu,diener,cheng}. 
They consider either the $F = 1$ spinor BEC 
by taking into account the d-d interaction or $F=3$ for
$^{52}$Cr atom gases in a realistic situation.  Here one must handle a 
7-component spinor with 5 different interaction channels 
$g_0$, $g_2$, $g_4$, $g_6$, and $g_{d}$.  The parameter 
space to hunt is large and difficult enough to find a stable configuration. 
The situation becomes further hard towards a larger $F$ 
where the d-d interaction is more important and eventually
dominant one among various channels.

Here we investigate generic properties of the spinor dipolar BEC 
under an optical trapping where the d-d interaction 
dominates other interactions except for the $s$-wave
repulsive channel. A proposed model Hamiltonian 
is intended to capture essential properties of the spinor BEC system. 
We note that this long-ranged and anisotropic d-d 
interaction has fascinated researches for a long time, 
for example, Luttinger and Tisza in their seminal paper \cite{luttinger} theoretically discussed 
the stable spin configurations of a spin model on a lattice
where classical spins with a fixed magnitude are free to rotate on a lattice. 
The present paper is designed to generalize this lattice spin model to a dipolar BEC system.
Here we are interested in the interplay  between the spin degrees of freedom and
the mass density through the d-d interaction.

We approach this problem from atomic species with large magnetic moment. 
This spinor dipole BEC with the hyper fine state $F$ ($F_z = -F, -F + 1, \cdots, F$) 
is characterized by $2F +1$ components $\Psi_\alpha ({\bf r})$. 
In general the number of the interaction channels are $F + 1$. 
For example, the $F = 1$  spinor BEC \cite{ohmi,ho}  is 
characterized by the scattering lengths  $a_0$ 
and $a_2$,  leading to the spin independent repulsive 
interaction $g_0 = 4 \pi \hbar^2 (a_0 + 2 a_0) / 3m$ 
and the spin dependent exchange interaction 
$g_2 = 4 \pi \hbar^2 (a_2 - a_0) /3m$. 
Since $a_0$ and $a_2$ are comparable, $g_2$ is actually small; 
$|g_2|/g_0\sim 1/10$ for $^{23}$Na \cite{stenger, burke} and $\sim 1/35$ 
for $^{87}$Rb \cite{barrett, klausen}. 
This tendency that, except for the dominant repulsive part $g_0$,
other spin-dependent channels are nearly cancelled 
is likely to be correct for other $F$'s \cite{hirano}. 

We can take a view in this paper that instead 
of working with $\vec{\Psi} ({\bf r})$ full quantum mechanical
$2F + 1$ components $( \Psi_{F}, \Psi_{F-1}, \cdots, \Psi_{-F})$ 
with the interaction parameters $g_0$, 
$g_2$, $g_4$, ..., and $g_{2F}$, the order parameter 
can be simplified to $\vec{\Psi} ({\bf r}_i) = 
\psi({\bf r}_i) {\vec S}({\bf r}_i)$ 
where ${\vec S}({\bf r}_i)$ is a classical vector with $| {\vec S} ({\bf r}_i)|^2 = 1$. 
Namely we can treat it as the classical spin vector whose 
magnitude $|\psi({\bf r}_i)|^2$ is proportional to the local condensate density.
In other words, we focus on long-wavelength and  low energy textured solutions of 
a dipolar system which will manifest the interplay between the mass and
spin density degrees of freedom.

We start with the following minimal model Hamiltonian

\begin{eqnarray}
H
&=& \int d^3{\bf r}_i \vec{\Psi}^\dagger ({\bf r}_i)
H_0 ({\bf r}_i)
\vec{\Psi} ({\bf r}_i) \nonumber \\
&+& \frac{1}{2} g_d \int \int d^3{\bf r}_i d^3{\bf r}_j
V_{dd} ({\bf r}_i, {\bf r}_j)
|\psi ({\bf r}_i)|^2 |\psi ({\bf r}_j)|^2,
\end{eqnarray}
\begin{eqnarray}
H_0
&=& - \frac{\hbar^2}{2 m} \nabla_i^2
+ V_{\rm trap} ({\bf r}_i)
- \mu + \frac{g}{2} | \vec{\Psi} ({\bf r}_i) |^2,
\end{eqnarray}
\begin{eqnarray}
V_{dd} ({\bf r}_i, {\bf r}_j)
&=& \frac{\vec{S}_i \cdot \vec{S}_j
- 3 (\vec{S}_i \cdot \vec{e}_{ij})
(\vec{S}_j \cdot \vec{e}_{ij})}
{r_{ij}^3},
\end{eqnarray}
where  $\vec{e}_{ij} \equiv ({\bf r}_i - {\bf r}_j) / r_{ij}$ with $r_{ij}=|{\bf r}_i - {\bf r}_j|$.
The uniaxially symmetric trap potential is given by $V_{\rm trap} ({\bf r}) = \frac{1}{2} m \omega^2 \{\gamma (x^2 + y^2)+z^2\}$ with $\gamma$ being the anisotropy parameter. $\mu$ is the chemical potential. The repulsive ($g > 0$) and the dipole-dipole  ($g_d$) interaction are introduced. The classical spin vector $\vec{S}_i\equiv\vec{S} ({\bf r}_i)$ characterizes the internal degrees of freedom of the system at the site $i$ and is denoted by spherical coordinates ($\varphi({\bf r}_i)$, $\theta({\bf r}_i)$) with $|\vec{S}_i|^2 = 1$. A dimensionless form of this Hamiltonian may be written as

\begin{widetext}
\begin{eqnarray}
H
&=& {1\over 2}\int d^3{\bf r}_i \Biggl [ | \nabla \psi ({\bf r}_i) |^2
+ n_i
\Bigl \{ \bigl (\nabla \theta ({\bf r}_i) \bigr )^2
+ \sin^2 \theta ({\bf r}_i)
\bigl ( \nabla \varphi ({\bf r}_i) \bigr)^2
+\gamma^2 (x^2 + y^2) + z^2\Bigr \} - 2 \mu n_i
+ g n_i^2
\Biggr ] \nonumber \\
&& \mbox{} + \frac{g_d}{2} \int \int d^3{\bf r}_i d^3{\bf r}_j
\frac{1}{r_{ij}^3} \left \{
\vec{S}_i \cdot \vec{S}_j
- 3 (\vec{S}_i \cdot \vec{e}_{ij})
(\vec{S}_j \cdot \vec{e}_{ij})
 \right \}
n_i n_j,
\end{eqnarray}
\end{widetext}
with $|\psi ({\bf r}_i)|^2 = n_i$. We note that the spin gradient term  in the first line is a Non-linear sigma model\cite{haldane}. Here it is extended to include the dipole-dipole interaction between the different parts of the spin density. The energy (length) is measured by the harmonic frequency $\omega$ (harmonic length $d\equiv 1/\sqrt {m\omega}$) with $\hbar = 1$ The functional derivatives with respect to $\psi^\ast ({\bf r}_i)$, $\varphi ({\bf r}_i)$ and $\theta ({\bf r}_i)$ lead to the corresponding Gross-Pitaevskii equations.

In this paper under a fixed repulsive interaction ($g / \omega d^3 = 0.01$) we vary the d-d interaction $g_d$ in a range of $0 \leq g_d \leq 0.4g$, beyond which the system is unstable. We consider two types of the confinement: A pancake ($\gamma = 0.2$) and a cigar ($\gamma = 5.0$) to see the shape dependence of the d-d interaction, which is long-ranged and anisotropic. The total particle number $\sim 10^4$. The three dimensional space is discretized into the lattice sites  $\sim 2.5\times 10^4$. Using the imaginary time ($\tau$) evolution of Gross-Pitaevskii  equations {\it e.g.} $\partial \psi_i/\partial \tau = -\delta H/\delta \psi^{\ast}_i$, we obtain stable configurations for spin and particle densities by starting with various initial patterns.

\begin{figure}[h!]
\centerline{\includegraphics[width=\linewidth]{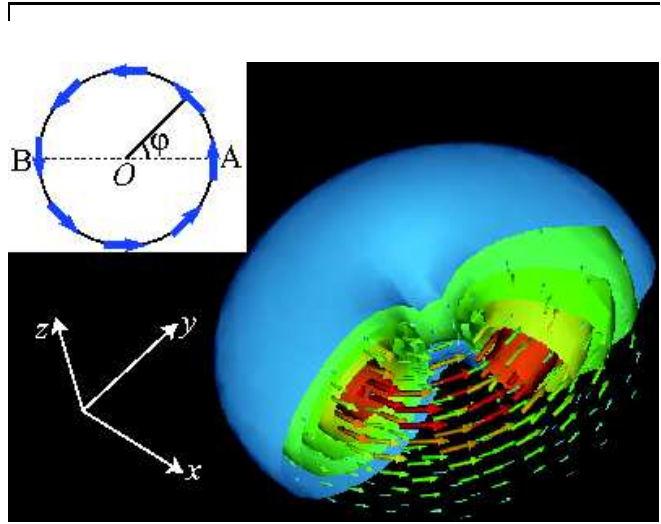}}
\caption{
Stereographic view of the spin current texture, displaying  simultaneously the number and spin densities. The pancake ($\gamma=0.2$)
is distorted and at the center the number density is depleted to give a doughnut like shape. $g_d=0.2g$. All spins lie in the $x$-$y$ plane, {\it i.e.} a coplanar spin structure, circulating around the origin $O$. The length of the arrow
is proportional to its number density.
Inset shows the schematic spin configuration on $z=0$ plane.
}
\label{fig:spincurrent}
\end{figure}

We start with the pancake shape ($\gamma = 0.2$). 
Figure \ref{fig:spincurrent} shows a stereographic image of the particle density and spin distributions. We call it spin current texture, where the spin direction circulates around the origin and is confined into the $x$-$y$ plane without the third component, that is, a coplanar texture. It is seen that the particle density distribution is strongly coupled to the spin one; In the central region the particles are depleted over the coherent length $\xi_d$ of the d-d interaction. In the present case $\xi_d\sim 2.0\xi_c$ ($\xi_c$ is the
ordinary coherent length of the repulsive interaction).

This spin current texture can be readily explained in the following way: 
 (1) Locally, along the stream line 
of the spin current the head-to-tail configuration minimizes the energy. (2) Globally, the spins at  A and B which are situated far apart about the origin $O$ shown in inset of Fig. \ref{fig:spincurrent} are orientated anti-parallel to minimize the d-d interaction. (3) When the two antiparallel spins at A and B come closer towards the origin $O$, the kinetic energy due to the spin modulation increases. To avoid this energy loss, the particle number is depleted in the central region at the cost of the harmonic potential energy.

For an alternative explanation of the spin current texture we rewrite the d-d interaction as $v_{dd} ({\bf r}_{ij}) \propto \frac{g_d}{r_{ij}^3} \sum_{\mu = -2}^2 Y_{2 \mu} (\cos \theta) \Sigma_\mu (ij)$ with $\Sigma_\mu (ij)$ being a rank 2 tensor consisting of the two spins at $i$ and $j$ sites, and $Y_{2 \mu} (\cos \theta)$ a spherical harmonics \cite{pethick}. $\theta$ is the polar angle in spherical coordinates of the system. The spin current texture shown in inset of Fig. \ref{fig:spincurrent} picks up the phase factor ${\rm e}^{2 i \varphi}$ when winding around the origin. This is coupled to $Y_{2 \pm 2} (\cos \theta) \propto \sin^2 \theta$, meaning that this orbital moment dictates the number density depletion at the pancake center. The spin-orbit coupling directly manifests itself here. The total angular momentum consisting of the spin and orbit ones is a conserved quantity of the present axis-symmetric system, leading to the Einstein-de Haas effect \cite{kawaguchi}. The spin current texture is stable for the wide range of anisotropy $\gamma$: $0.01 \leq \gamma \leq 0.6$, beyond which it becomes unstable.

\begin{figure}
\centerline{\includegraphics[width=\linewidth]{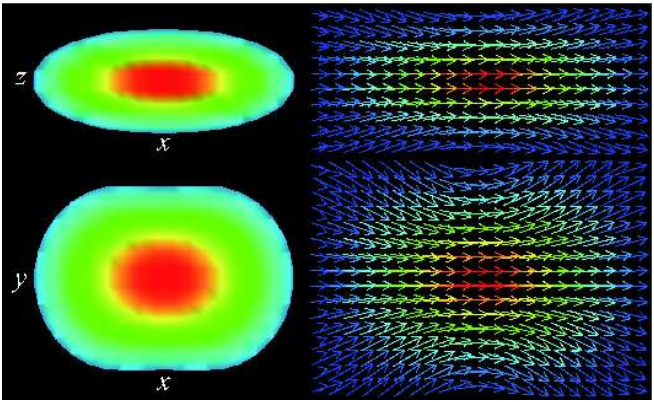}}
\caption{
The r-flare texture. Left (right) column shows the cross-sectional density plots of the particle number (the corresponding spin structure). The circular profile in the $x$-$y$ plane is spontaneously broken. $g_d=0.2g, \gamma=0.2.$}
\label{fig:rflare}
\end{figure}

\begin{figure}
\centerline{\includegraphics[width=\linewidth]{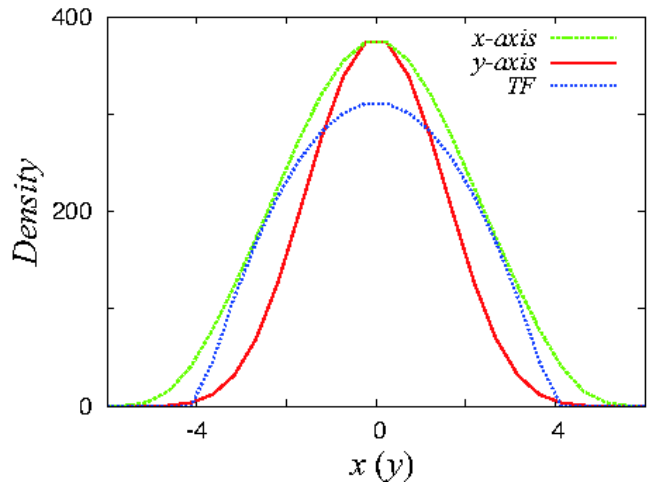}}
\caption{
Cross sections of the particle number in Fig. \ref{fig:rflare} along the $x$ and $y$-axis
compared with Thomas-Fermi (TF) profile for $g_d=0$. 
The profile is elongated (compressed) along
the $x$ ($y$)-axis.}
\label{fig:density}
\end{figure}

Figure \ref{fig:rflare} displays another stable configuration in a similar situation.
The left (right) column shows the density plots of the particle number (the corresponding spin structure). The spins are almost parallel to the $x$-axis, but at the outer region they bent away. We call it r-flare texture, which is a non-coplanar spin arrangement. It is clearly seen that the axis-symmetry in the $x$-$y$ plane, which was originally circular,
is spontaneously broken so that the circular shape is elongated
along the $x$-axis and compressed along the $y$-axis. Figure \ref{fig:density} displays the $x$ and $y$-axis cross-sections of the particle density, compared with the Thomas-Fermi (TF) profile for $g_d = 0$ with the same particle number. Because of the d-d interaction which favors the head-to-tail arrangement, the particle number is increased at the center. The bending tendency at the circumference increases with increasing $g_d$. Beyond a certain critical value $g_d \cong 0.27 g$ for $\sim 10^4$ particles, the r-flare texture becomes unstable, indicating a quantum phase transition.
Upon increasing the total particle number the r-flare is replaced by the 
spin current texture.
We also note that the z-flare texture
in which the polarization points 
to the $z$-axis is equally stable as we explain shortly.

\begin{figure}
\centerline{\includegraphics[width=\linewidth]{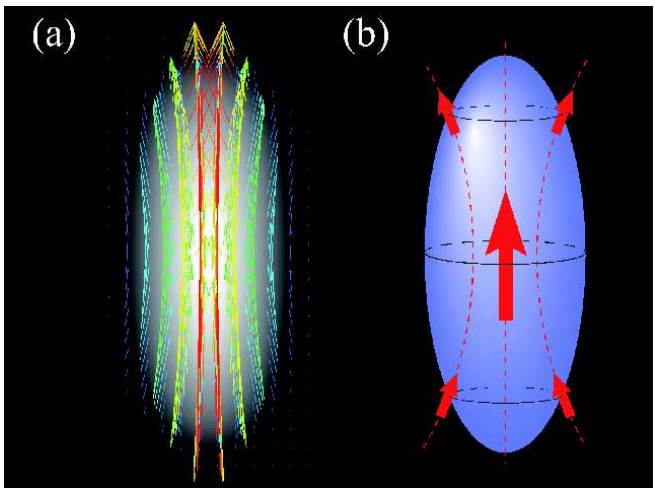}}
\caption{
(a) The z-flare spin texture in the cigar trap along the $z$-axis.
The spins almost point to the $z$ direction. In the outer regions
they bent. The bright region in background corresponds to high number density.
$g_d=0.2g, \gamma=5.0.$
(b) Schematic figure to explain this spin configuration
due to d-d interaction.}
\label{fig:zflare}
\end{figure}

Let us turn to the cigar shape case  elongated along the $z$-axis 
with the trap anisotropy $\gamma=5.0$.
The stable configuration we obtain is shown in Fig. \ref{fig:zflare} where
the spin structure is basically a flare spin texture which is a non-coplanar spin arrangement. 
Namely, the bending occurs radially so that the spin texture is a
three dimensional object, but keeps axis-symmetry around the $z$-axis. 
The particle density is modified from the TF profile for $g_d=0$, elongated along the 
$z$ direction and compressed to the $z$-axis.

\begin{figure}
\centerline{\includegraphics[width=\linewidth]{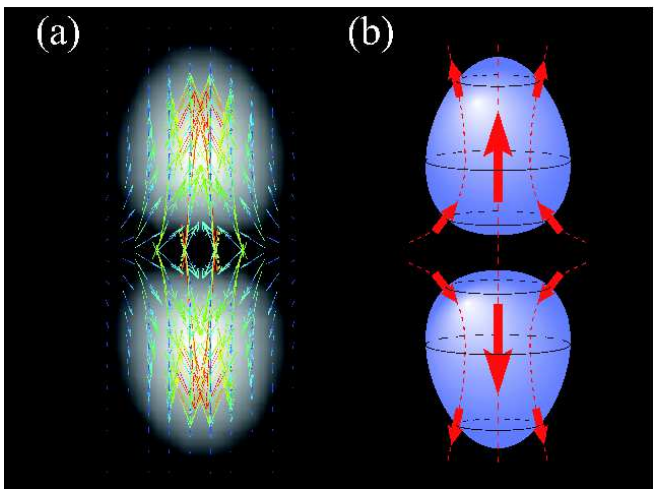}}
\caption{
(a) The two-z-flare spin texture 
under the same parameter set ($g_d=0.2g, \gamma=5.0.$)
as in Fig. \ref{fig:zflare} with different initial spin configuration.
The bright region in background corresponds to high number density.
(b) Schematic figure to explain this spin structure. At the 
$z=0$ plane two oppositely aligned spins meet and the number density 
is depleted.
}
\label{fig:2zflare}
\end{figure}

This can be understood by seeing Fig. \ref{fig:zflare} (b). The up-spin density near the center
exerts the d-d force so as to align the outer spins parallel to the vector connecting
the center and its position, taking the head-to-tail configuration. This results in a non-coplanar structure, but the axis-symmetry about the $z$-axis is preserved.
This spin texture is stable for $g_d\leq 0.3g$ and robust for different aspect ratios:
$\gamma$ = 0.2 and 1.5. The bending angle of the flare spin texture
increases and the elongation along the $z$ direction becomes larger
as $g_d$ increases (= 0.1 and 0.2).

Finally we display an example to show how the model Hamiltonian admits 
many subtle spin textures with comparable energies.
Figure \ref{fig:2zflare} (a) shows  the two-z-flares oppositely polarized stacked back to back.
This configuration is stabilized starting with a hedgehog 
spin configuration, or skyrmion at the center from which all the spins point outward from the origin.
In the end the two-z-flares oppositely polarized become stable, but at the central $z=0$  plane the antiparallel spins meet as seen from Fig. \ref{fig:2zflare} (b). To avoid drastic changes of the spin direction, or the spin kinetic energy loss,
the particle density decreases there. As a result even though the harmonic potential 
energy is minimal there, the two-z-flare spin textures oppositely polarized
are stacked back to back, but two objects are almost split.
This example illustrates strong coupling  between the particle number and spin densities through the d-d interaction.

These spin textures can be observed directly via a
novel phase-sensitive {\it in situ} detection \cite{review}
or indirectly via conventional absorption imaging for the number density.
It is interesting to examine the vortex properties under rotation.
For the spin current texture, the vortex entry into a system
should be easy because in the central region the mass density
is already depleted. We point out that the collective modes might 
be also intriguing because the mass density is tightly coupled with
the spin degrees of freedom. These problems belong to future work.

In summary, we have introduced a model Hamiltonian
to capture the essential nature of dipolar spinor
BEC where the spin magnitude is large enough, focusing on long wavelength and
low energy textured solutions. 
We show several typical stable configurations by solving 
the Gross-Pitaevshii equation where the spin and mass densities
are strongly coupled due to the dipole-dipole interaction.
The shape of the harmonic potential trapping is crucial to 
determine the spin texture. The model Hamiltonian is a minimal extension of the Non-linear sigma model with the d-d interaction, and yet complicate 
and versatile enough to explore further because it is expected that there
are many stable configurations with comparable energies. Finally the model Hamiltonian
is applicable literally for electric dipolar systems without further 
approximation.  We expect that BEC of hetero-nuclear molecules
with permanent electric dipole moment might be realized in near future \cite{molecule} where the formation of such textures may be possible.

We thank Tarun K. Ghosh and W. Pogosov for useful discussions in the early stage of this research. This work of S. G. was supported by a grant of the Japan Society for the Promotion of Science.

\end{document}